\documentclass[prb,showpacs,amsfonts,superscriptaddress,twocolumn]{revtex4}%
\usepackage{amsmath}
\usepackage{amsfonts}
\usepackage{amssymb}
\usepackage{graphicx}%
\begin{document}
\title{Effects of disorder on the non-zero temperature Mott Transition}
\author{M. C. O. Aguiar}
\affiliation{Center for Materials Theory, Serin Physics Laboratory, Rutgers University, 136
Frelinghuysen Road, Piscataway, New Jersey 08854}
\author{V. Dobrosavljevi\'{c}}
\affiliation{Department of Physics and National High Magnetic Field Laboratory, Florida
State University, Tallahassee, FL 32306}
\author{E. Abrahams}
\affiliation{Center for Materials Theory, Serin Physics Laboratory, Rutgers University, 136
Frelinghuysen Road, Piscataway, New Jersey 08854}
\author{G. Kotliar}
\affiliation{Center for Materials Theory, Serin Physics Laboratory, Rutgers University, 136
Frelinghuysen Road, Piscataway, New Jersey 08854}


\begin{abstract}
The physics of the metal-insulator coexistence region near the non-zero
temperature Mott transition is investigated in presence of weak disorder. We
demonstrate that disorder reduces the temperature extent and the general size
of the coexistence region, consistent with recent experiments on several Mott
systems. We also discuss the qualitative scenario for the disorder-modified
Mott transition, and present simple scaling arguments that reveal the
similarities to, and the differences from, the clean limit.

\end{abstract}
\maketitle

\section{Introduction and motivation}
The physics of the metal-insulator transition has continued to attract
considerable interest in recent years. Substantial progress has been achieved
in understanding the behavior near the interaction-driven transition, where
Dynamical Mean Field Theory~\cite{georges} (DMFT) has been very
successful in explaining the behavior of several classes of materials ranging
from transition metal oxides such as V$_{2}$O$_{3}$ to organic Mott systems.
This approach has been especially useful in describing the non-zero temperature
behavior in the paramagnetic coexistence region between
the metal and the insulator. In this regime, the two phases compete, and the
resulting behavior emerges as a compromise between the energy gain to form
coherent quasiparticles, and the larger entropy inherent to the incoherent
insulating solution. There is not actual two-phase coexistence (as in conventional first order thermodynamic phase transitions) in this region. Rather, it is a region of parameters in which two local minima of the free energy coexist.

So far, most theoretical work has concentrated on clean systems, although
several experimental studies  indicate that effects of disorder are
particularly important precisely in this coexistence regime. Measurements
performed in compounds such as NiSSe mixtures~\cite{sekine,matsuura,miyasaka}
and $\kappa$-organics~\cite{limelette,strack}  indicate that the
presence of disorder pushes down the critical temperature end point of the
metal and insulator coexistence region. In particular, experiments performed
on a NiS$_{2}$ compound, which has much weaker disorder, show that the Mott
transition occurs at 150K,~\cite{sekine} with an external applied pressure of 3
GPa, while in the substituted NiS$_{2-x}$Se$_{x}$ compound it is seen only
below 100K.~\cite{miyasaka} It is important to notice that applying an
external pressure to these compounds is equivalent to substituting S by Se,
which might suggest that the results above would be in conflict. A speculation
was made that the reduction in the transition temperature would be due to the
local randomness introduced with Se substitution.\cite{matsuura}

We address the theoretical issues from the perspective of the Hubbard model. It is not a priori obvious what should be the effect of disorder on the size and the temperature range of the coexistence region. On the one hand, disorder tends to broaden the Hubbard bands and thus larger interaction is needed to open a Mott Hubbard gap. This
may lead to a larger overall energy scale, which could stabilize the
coexistence region. On the other hand, disorder generally leads to spatial
fluctuations in all local quantities, an effect that could smear or decrease
the jump at any first order phase transition, and thus reduce the coexistence
energy scale. These considerations indicate that careful theoretical work is
called for, which can address the interplay of interactions and disorder near
the Mott metal-insulator transition.

A formalism that describes the effects of disorder within a DMFT approach was
outlined some time ago,\cite{dmftdis} but a very limited number of
calculations were explicitly carried out within this framework. More recently,
the approach was reexamined to investigate strong correlation effects on
disorder screening,\cite{screening} and the related temperature dependence of
transport in the metallic phase.\cite{inelastic} These results shed light on
several puzzling phenomena observed in experiments on two dimensional electron
systems, but did not provide a description of the physics relevant to the
coexistence region at non-zero temperature.

In this paper we examine the phase diagram for the Mott transition in the
presence of moderate disorder at non-zero temperature within the DMFT
approach.\cite{dmftdis} We present results describing the evolution of the
coexistence region, showing that disorder generally reduces its size, in
agreement with experiments. Our results give a physical picture that
describes the gradual destruction of quasiparticles as the Mott insulator is
approached, and establish the qualitative modification of the critical behavior
resulting from the presence of disorder. 

Our findings are valid in the
regime of strong correlations but weak to moderate disorder, where Anderson
localization effects, which are neglected in our theory, can be
safely ignored. The latter have been included in earlier zero temperature DMFT-based strong correlation calculations. \cite{statdmft,vollhardt} In particular, we mention that our lowest temperature results are consistent with the $T=0$ result at weak disorder of Byczuk et al,\cite{vollhardt} but give the temperature dependence of the metal-insulator coexistence region.

\section{non-zero temperature DMFT for disordered electrons}

We consider a half-filled Hubbard model in the presence of random site
energies, as given by the hamiltonian%

\begin{equation}
H=-t\sum_{<ij>\sigma}c_{i\sigma}^{\dagger}c_{j\sigma}+\sum_{i\sigma
}\varepsilon_{i}n_{i\sigma}+U\sum_{i}n_{i\uparrow}n_{i\downarrow}.
\end{equation}

Here $c_{i\sigma}^{\dagger}$ ($c_{i\sigma}$) creates (destroys) a conduction
electron with spin $\sigma$ on site $i$, $n_{i\sigma}=c_{i\sigma}^{\dagger
}c_{i\sigma}$ is the particle number operator, $t$ is the hopping amplitude,
and $U$ is the on-site repulsion. The random site energies $\varepsilon_{i}$
are assumed to have a uniform distribution of width $W$.

Within DMFT for disordered electrons, \cite{dmftdis} a quasiparticle is
characterized by a local but site-dependent \cite{zimanyi} self-energy
function $\Sigma_{i}(\omega)=\Sigma(\omega,\varepsilon_{i})$. To calculate
these self-energies, the problem is mapped onto an \textit{ensemble} of
Anderson impurity problems~\cite{dmftdis} embedded in a self-consistently
calculated conduction bath. In this approach, only quantitative details of the
solution depend on the details of the electronic band structure; in the
following we concentrate on a semi-circular model density of states. In this
particular case, the hybridization function is given by
\begin{equation}
\Delta(\omega)=t^{2}\bar{G}(\omega) \label{deltahd}%
\end{equation}
\noindent and the average local Green's function, $\bar{G}(\omega)$, is
obtained by imposing the following self-consistent condition%
\begin{equation}
\bar{G}(\omega)=\left\langle \frac{1}{\omega-\varepsilon_{i}-\Delta
(\omega)-\Sigma_{i}(\omega)}\right\rangle , \label{autoconh}%
\end{equation}
\noindent where $\left<  ...\right>  $ indicates the arithmetic average over
the distribution of $\varepsilon_{i}$.

To solve the single-impurity problems at non-zero temperature for different site
energies, we mostly used the iterated perturbation theory (IPT) method of
Kajueter and Kotliar.\cite{kajueter,nolting} However, to check the accuracy of
the results, in several instances we also used the numerically exact quantum
Monte Carlo method as an impurity solver, and generally found good qualitative
and even quantitative agreement, supporting the validity of our IPT
predictions in the relevant parameter ranges. Throughout the paper we express all energies in
units of the bandwidth.

\section{Phase diagram}

\begin{figure}[ptb]
\begin{center}
\includegraphics[
trim=0.000000in 0.000000in 0.000000in -1.1in,
height=2.2in, width=2.6in ]{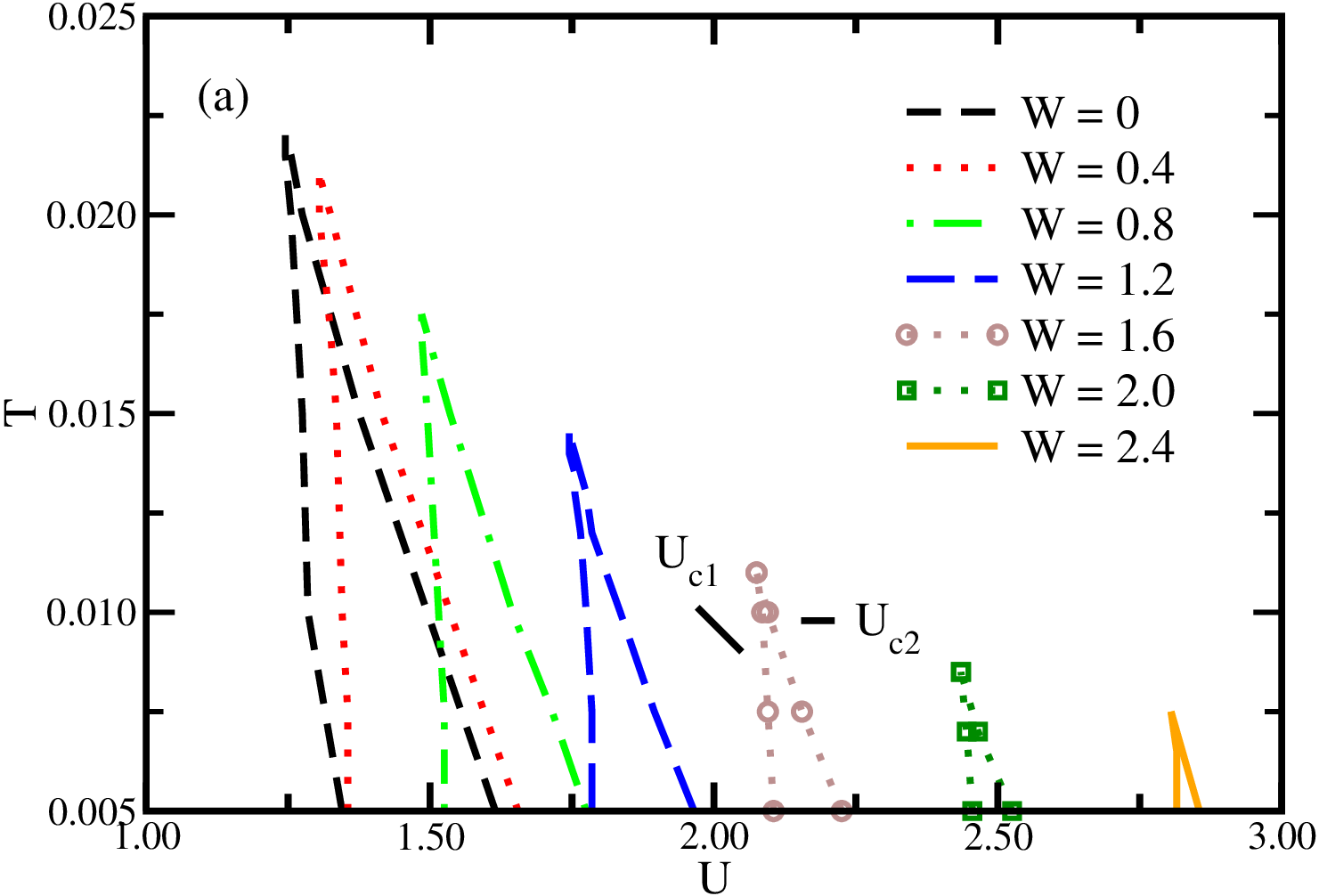}
\includegraphics[
trim=0.000000in 0.000000in 0.000000in -1.1in,
height=2.2in, width=2.6in ]{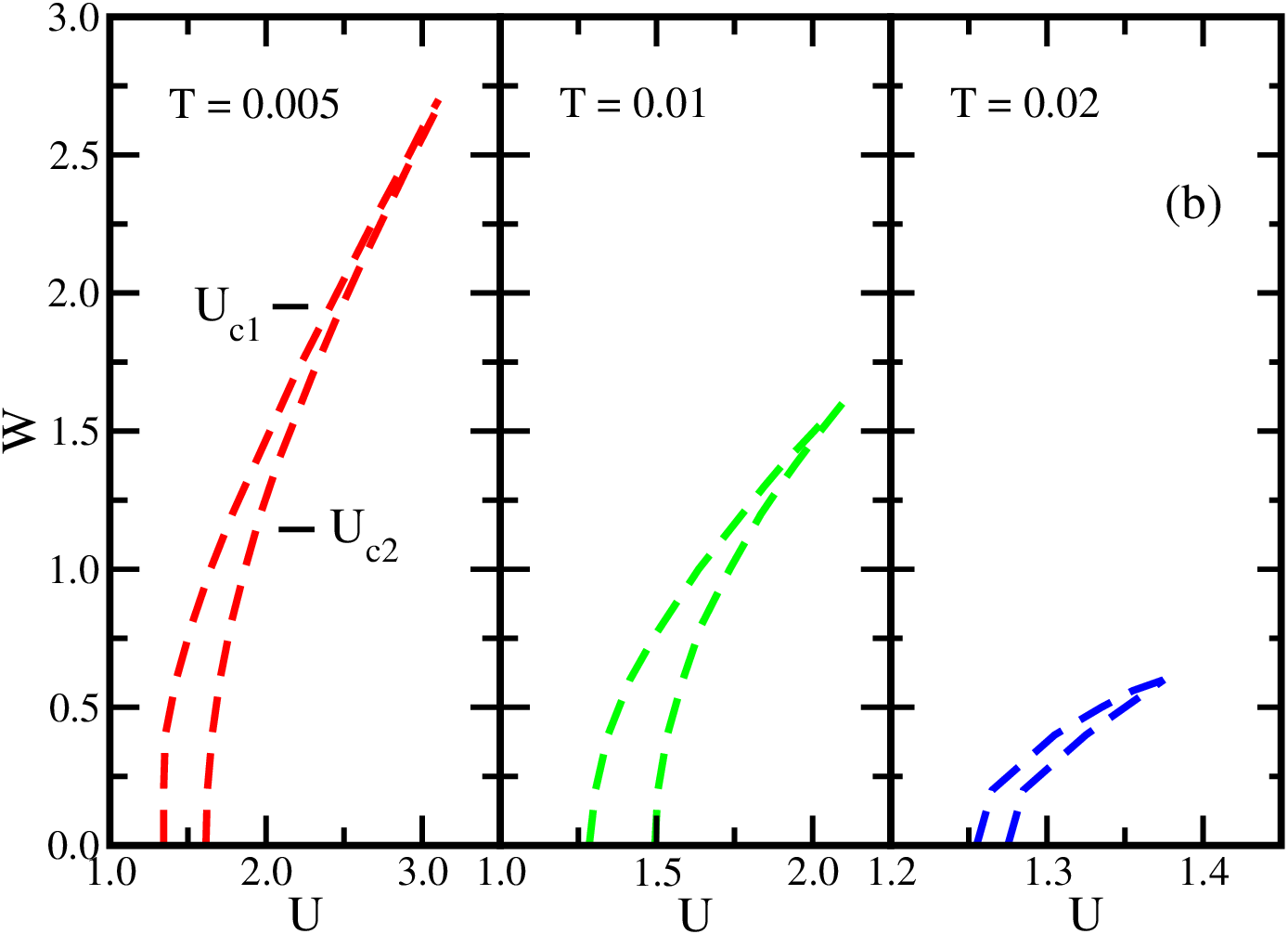}
\end{center}
\caption{Phase diagram for the disordered Hubbard model at non-zero temperature.
(a) ($U$,$T$) diagram for different disorder strengths. (b) ($U$,$W$) diagram
at different temperatures. $U_{c1}$ and $U_{c2}$ lines are indicated in one of
the plots, but similar definitions apply to the other results as well.}%
\label{fig1}%
\end{figure}

We first examine the evolution of the coexistence region as disorder is
introduced. Within this region, both metallic and  insulating
solutions are found, depending on the initial guess used in the iterative
scheme for solving the self-consistency condition. Typical results are
presented in Fig.~\ref{fig1}, showing the phase diagram obtained within
DMFT-IPT at non-zero temperature, for varying levels of disorder $W$. For each
level of disorder [shown in panel (a)] or temperature [shown in panel (b)],
the first (from left) of the two lines, the so-called $U_{c1}$, indicates the
stability boundary (i.e. the spinodal) of the insulating solution. Conversely,
the second of the two lines, identified as $U_{c2}$, represents the boundary
of the metallic solution. The coexistence region is found between these two
lines, i.e. for $U_{c1}<U<U_{c2}$. Our results are in good quantitative
agreement with previous results obtained in the $T=0$ limit in presence of
disorder,\cite{screening} and also with non-zero temperature results in absence
of disorder.\cite{georges} As the disorder increases, the metal-insulator transition
generally moves to larger $U$. Physically, this reflects the fact that
disorder broadens the bands and smears the gap, making it harder for the
Mott-Hubbard gap to open, so that a larger $U$ is necessary for the
transition. At the same time, the temperature-dependent coexistence region is
found to shrink [Fig.~\ref{fig1}(a)], persisting only below a critical
end-point temperature $T_{c}(W)$ . At any given temperature, the principal
effects of introducing disorder [Fig.~\ref{fig1}(b)] are as follows: (1) both
the $U_{c1}$ and $U_{c2}$ lines move towards larger interaction potential; (2)
the lines become closer to each other as disorder increases. In fact, they
both approach the $W=U$ line as $W\rightarrow\infty$.

Having obtained these results in quantitative detail, we would like to
understand the physical origin of this behavior. In the following we present
simple analytical arguments relating the non-zero temperature aspects of the
coexistence region to the evolution of its ground state properties. Our
strategy is motivated by the following observations: (a) the shape of the
non-zero temperature coexistence region [Fig.~\ref{fig1}(a)] remains
\textit{very similar} at different values of disorder; (b) its size, both in
terms of temperature and in terms of $U$-range, shrinks as disorder increases.
This suggests that the physical mechanism for the destruction of the
coexistence region as the temperature  increases is similar to that of the clean
limit, where it is governed by decoherence processes due to inelastic
electron-electron scattering. Therefore, we begin our analysis by concentrating on the
clean limit, where we show how simple estimates for the critical end-point
temperature $T_{c}$ can be obtained.
\section{Coexistence region in the clean limit}
The coexistence region at non-zero temperature is delimited by the two spinodal
lines $U_{c1}(T)$ and $U_{c2}(T)$; the critical end-point temperature $T_{c}$
is reached when these two boundaries intersect. To estimate $T_{c}$ using the
$T=0$ properties of the model, we need to understand the temperature
dependence of each of these lines.
\subsection{Insulating spinodal}
The insulating spinodal $U_{c1}(T)$ essentially corresponds to the closing of
the gap separating the two Hubbard bands in the Mott insulator. Its
temperature dependence should thus reflect that of the Hubbard bands. In
contrast to the correlated metallic state close to the Mott transition, the
insulating solution is not characterized by a small energy scale in the
coexistence region. Accordingly, it is not expected to have strong temperature dependence; its weak temperature dependence reflects activated processes across the Mott-Hubbard gap. Such activations
only lead to (exponentially) weak rounding/broadening of the Hubbard bands,
which should very slowly reduce $U_{c1}(T)$ as temperature increases. Such
behavior is indeed clearly seen in our results. This temperature dependence
is, however, much weaker than that characterizing $U_{c2}(T)$. For purposes of
roughly estimating $T_{c}$, to leading order we can ignore this weak
temperature dependence, so that
\begin{equation}
U_{c1}(T)\approx U_{c1}(T=0).
\end{equation}

\subsection{Metallic spinodal}

In the vicinity of the Mott transition, the metallic solution is characterized
by a low energy scale corresponding to the coherence temperature $T^{\ast}$ of
a low-temperature Fermi liquid.\cite{georges}  Above $T^{\ast}$ the heavy quasiparticles are destroyed, and
the metallic solution becomes unstable. To estimate $U_{c2}(T)$ we need to
determine how this coherence temperature varies as the transition is
approached. From detailed studies of the clean\cite{georges} and
disordered\cite{inelastic} Hubbard models within DMFT, it is known that this
coherence temperature can be estimated as%
\begin{equation}
T^{\ast}\approx AT_{F}Z
\end{equation}
where $T_{F}$ is the Fermi temperature, $A$ is a constant of order one, and
$Z$ is the quasiparticle (QP) weight defined as%

\begin{equation}
Z=\left[  \left.  1-\frac{\partial}{\partial\omega}\mbox{Im}\Sigma
(\omega)\right\vert _{\omega\rightarrow0}\right]  ^{-1}. \label{e.2}%
\end{equation}
The behavior of $Z$ is well known in the clean
limit,\cite{georges} where it decreases linearly as $U$ increases toward the metallic spinodal, viz.
\begin{equation}
Z=C[U_{c2}(0)-U].
\end{equation}
From numerical studies,\cite{georges} the proportionality constant
$C\approx0.45$. Therefore, the coherence temperature can be written as%
\begin{equation}
T^{\ast}(U)=ACT_{F}[U_{c2}(0)-U].
\end{equation}

We can now  estimate the temperature dependence of $U_{c2}(T)$
as that value of the interaction needed to set $T^{\ast}(U)=T$, i.e.%
\[
T=ACT_{F}[U_{c2}(0)-U_{c2}(T)].
\]
In other words%
\begin{equation}
U_{c2}(T)\approx U_{c2}(0)-BT,
\end{equation}
where $B=1/ACT_{F}$. From our numerical results [see Fig.~\ref{fig1}(a)] we find
$B\approx22$, giving $A\approx0.2$, in reasonable agreement\cite{coeffA} with
estimates\cite{inelastic} from the literature.

Using these expressions for $U_{c1}(T)$ and $U_{c2}(T)$, we arrive at the
estimate for the critical end-point temperature%
\begin{equation}
T_{c}\approx\lbrack U_{c2}(0)-U_{c1}(0)]/B,
\end{equation}
which agrees within $10\%$ with our numerical results (see Fig.~\ref{fig7}).

\section{Critical behavior in presence of disorder}

Encouraged by the success of our analytical description of the coexistence
regime in the clean limit, we now turn our attention to the effects of
disorder. As in the clean limit, we would like to relate the finite
temperature properties to the critical behavior of the quasiparticles at
$T=0$. To do this, we therefore concentrate on describing the critical
behavior in presence of disorder.

\begin{figure}[ptb]
\begin{center}
\includegraphics[
trim=0.000000in 0.000000in 0.000000in -1.1in,
height=2.2in, width=2.6in ]{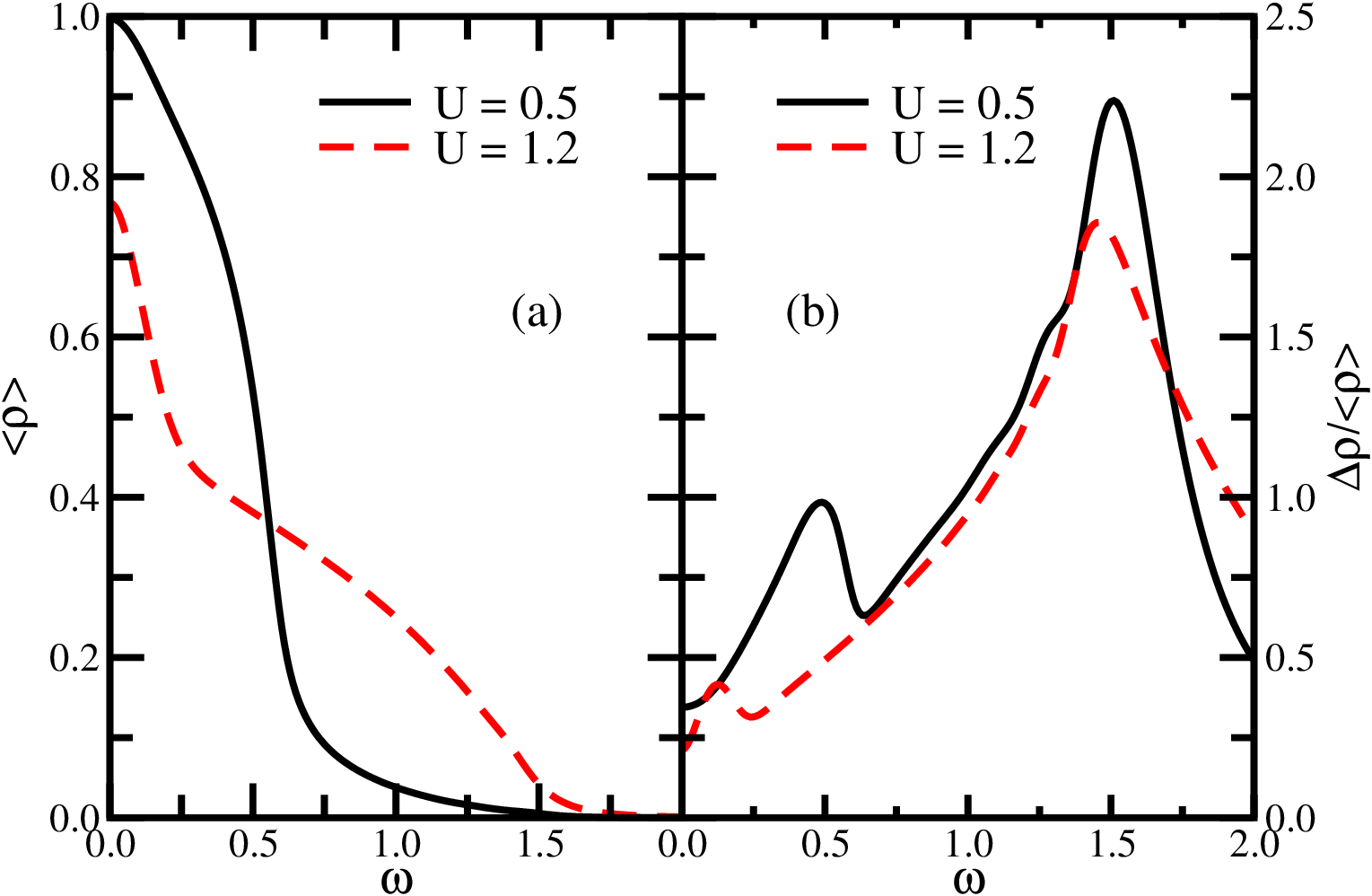}
\end{center}
\caption{(a) Average spectral function and (b)~relative deviation of the 
distribution of $\rho_i(\omega)$, $\Delta\rho/\left<\rho\right>$, as a 
function of frequency for different values of the interaction potential. 
$\Delta\rho$ is the standard deviation of the distribution of 
$\rho_i(\omega)$, which is given by $\sqrt{\sum_i(\rho_i(\omega)-\left<
\rho_i(\omega)\right>)^2/(N-1)}$, where $N$ is the number of local site 
energies considered. Other parameters used were $T=0.05$ and $W=1.0$.}%
\label{fig2}%
\end{figure}

The principal new feature introduced by disorder within the DMFT scheme is 
the spatial variation of the spectral function, $\rho_i(\omega)$. 
This is shown in Fig.~\ref{fig2} at all energy scales: on the left we 
have the average spectral function and on the right the relative deviation 
of its distribution, in the metallic phase. 
For each value of the interaction potential, the distribution of 
$\rho_i(\omega)$ presents a large dip at $\omega\approx0$ and becomes 
broader as the frequency increases. This comes from the fact that at small 
frequencies the system is in the Fermi liquid regime. At finite 
temperature we observe the reminiscence of the perfect disorder screening 
seen at $T=0$ close to the Mott transition.\cite{screening} For large 
frequencies, the quasiparticle regime is no more valid and the appropriate 
description is in terms of Hubbard bands, resulting in an increase of the 
fluctuation in $\rho_i(\omega)$. 

In the disordered case, the self-energy function $\Sigma_{i}(\omega)$ 
presents site-to-site fluctuations, which lead to the spatial variations 
of the spectral function discussed above. The QP weights 
$Z_{i}=Z(\varepsilon_{i})$ 
now depend on the local site energy $\varepsilon_{i}$. To properly describe 
the approach to the Mott transition, we therefore must follow the evolution 
of the entire function $Z(\varepsilon_{i})$ as the transition is 
approached.\cite{sitener}

\subsection{Behavior of local QP weights}

\begin{figure}[ptb]
\begin{center}
\includegraphics[
trim=0.000000in 0.000000in 0.000000in -1.1in,
height=2.2in, width=2.6in ]{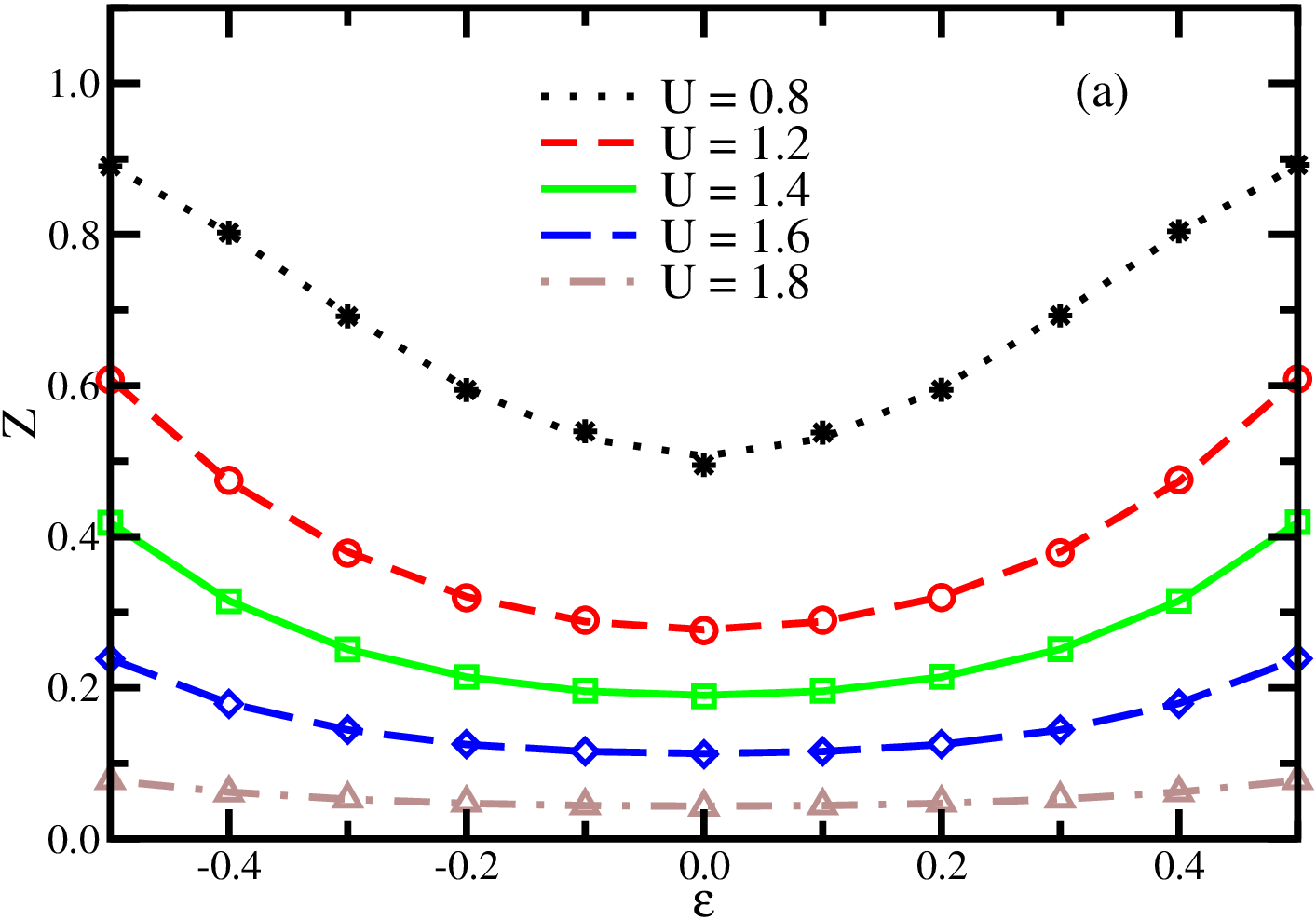}
\includegraphics[
trim=0.000000in 0.000000in 0.000000in -1.1in,
height=2.2in, width=2.6in ]{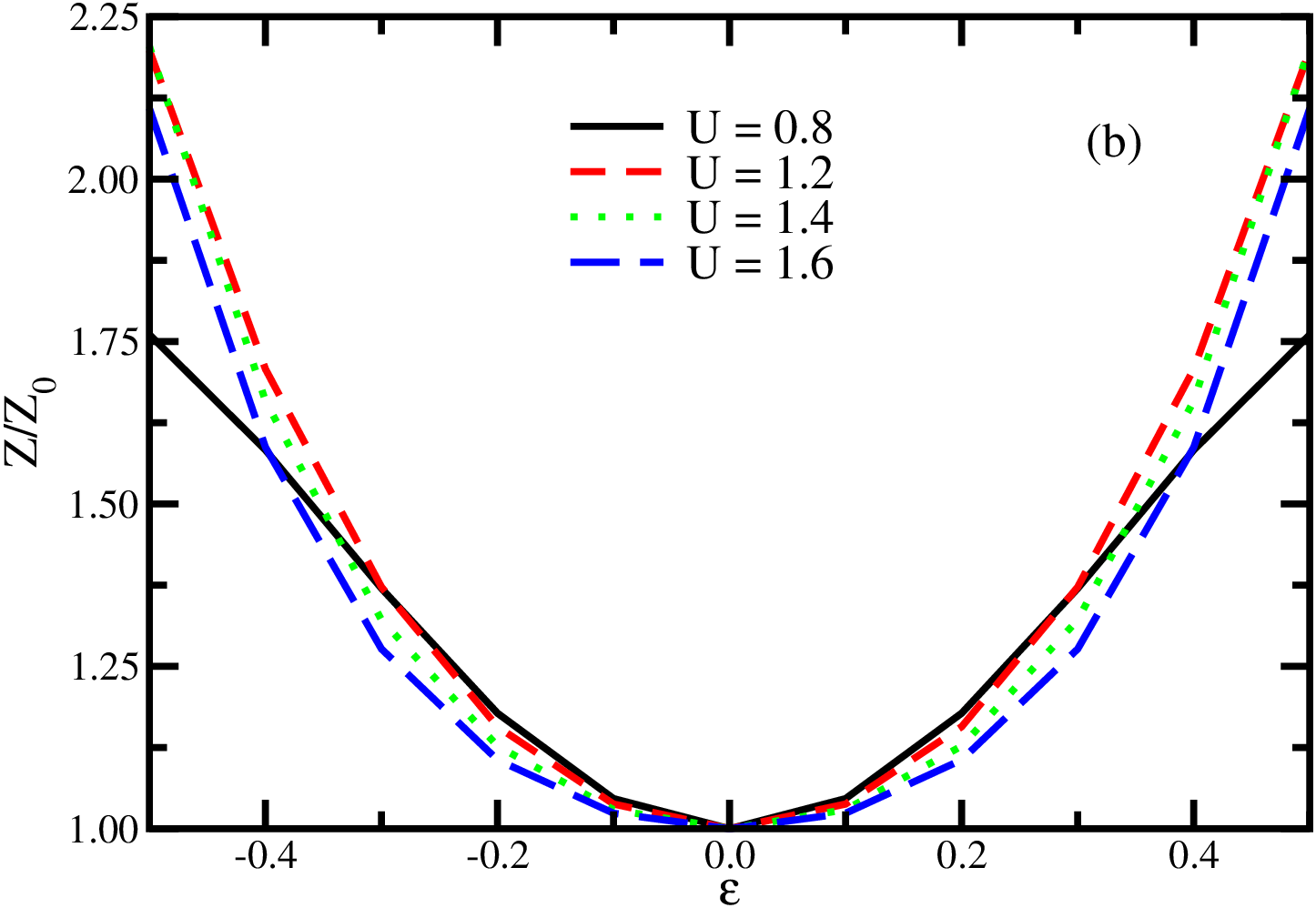}
\end{center}
\caption{(a) Quasiparticle weight as a function of the on-site energy for
different values of the interaction potential as the $U_{c2}$ line is
approached, for disorder strength $W=1.0$. The symbols are the numerical data,
while the lines correspond to the fitting indicated in the plot, that is to a
function with even exponents in $\varepsilon$. (b) Fitted results for $Z$
divided by $Z_{0}$ (the quasiparticle weight for $\varepsilon=0$) as a
function of $\varepsilon$, showing that close to the MIT the curves for
different $U$ scale. These results were obtained at a low but finite
temperature $T=0.005$.}%
\label{fig3}%
\end{figure}

Given the self-consistent solution of our ensemble of impurity models, we
calculate the local QP weights as

\begin{equation}
Z_{i}=\left[  \left.  1-\frac{\partial}{\partial\omega}\mbox{Im}\Sigma
_{i}(\omega)\right\vert _{\omega\rightarrow0}\right]  ^{-1}.
\end{equation}
Typical results are shown in Fig.~\ref{fig3}(a), where we plot $Z_{i}=Z(\varepsilon
_{i})$ at $T=0.005$, for disorder strength $W=1$, as the metallic spinodal is approached by
increasing the interaction $U$ toward $U_{c2}\approx 1.9$.
We first observe that for small $U$, away from the transition, the QP weights
$Z_{i}$ have strong $\varepsilon_{i}$ dependence, with the smallest $Z_{i}$
at $\varepsilon_{i}=0$. Physically, this reflects the tendency for
correlation effects (suppression of $Z$) to be the strongest on sites which
are locally close to half-filling (singly occupied). Nonzero site energies
favor the local occupation departing from half-filling, thus reducing the
correlation effect, and increasing $Z_{i}$.

As $U$ increases, all the $Z_{i}$'s decrease, as in the clean case. But how does this affect the distribution of
QP weights $Z_{i}=Z(\varepsilon_{i})$? At first glance it seems that the
$\varepsilon_{i}$ dependence becomes weaker, but a closer look reveals this
not to be the case. As we shall now demonstrate, all the $Z_{i}$'s decrease
linearly near the transition, i.e. they assume the form%
\begin{equation}
Z(U,\varepsilon_{i})=K(\varepsilon_{i})[U_{c2}-U],
\end{equation}
where only the prefactor $K(\varepsilon_{i})$ depends on $\varepsilon_{i}$.
If, to leading order, these prefactors remain independent of the distance to
the spinodal, then the entire family of curves $Z(U,\varepsilon_{i})$  can all be collapsed on a single
scaling function. To verify this hypothesis, we define reduced QP weights%
\begin{equation}
Z^{\ast}(\varepsilon_{i})=Z(U,\varepsilon_{i})/Z(U,0).
\end{equation}
If our scaling ansatz is valid, then the  $Z^{\ast
}(\varepsilon_{i})$ should approach a non-zero limit as $U\longrightarrow
U_{c2}$, i.e. they should all collapse onto a single scaling function.  As shown in Fig.~\ref{fig3}(b), this behavior is observed
only for $U$ sufficiently close to $U_{c2}$ [note that the data for $U=0.8$ (further
from the transition) show deviations from leading scaling]. This is precisely
what we expect, since such simple scaling behavior typically occurs only
within a critical region close to the metallic spinodal.

\subsection{Distribution $P(Z_{i})$ of local QP weights}

\begin{figure}[ptb]
\begin{center}
\includegraphics[
trim=0.000000in 0.000000in 0.000000in -1.1in,
height=2.2in, width=2.6in ]{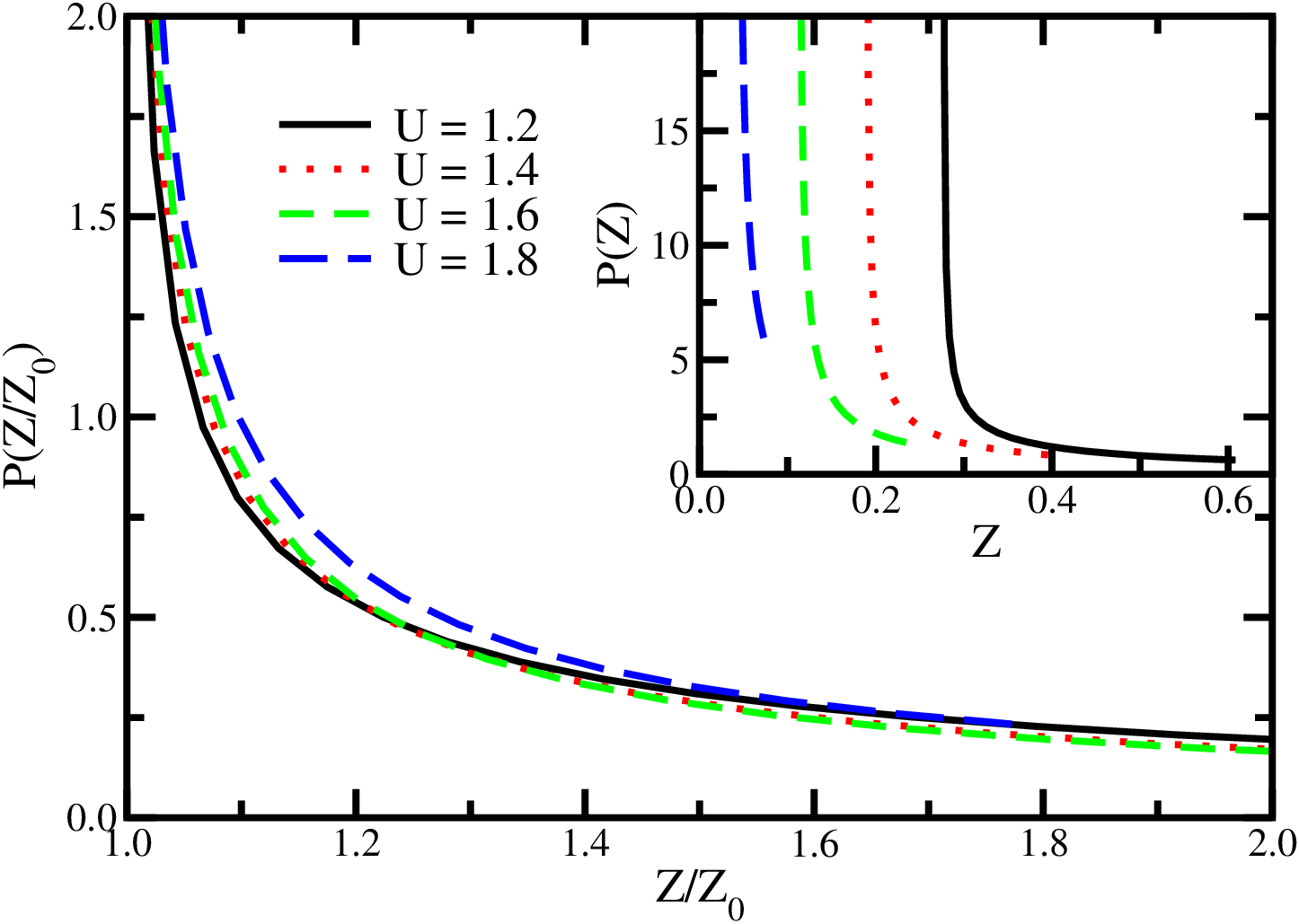}
\end{center}
\caption{Distribution of quasiparticle weight for different values of the
interaction potential. The main plot shows how the curves collapse when we
look at $Z/Z_{0}$. The inset shows the results for $Z$ itself. Other
parameters as in Fig.~\ref{fig3}.}%
\label{fig4}%
\end{figure}

\begin{figure}[ptb]
\begin{center}
\includegraphics[
trim=0.000000in 0.000000in 0.000000in -1.1in,
height=2.2in, width=2.6in ]{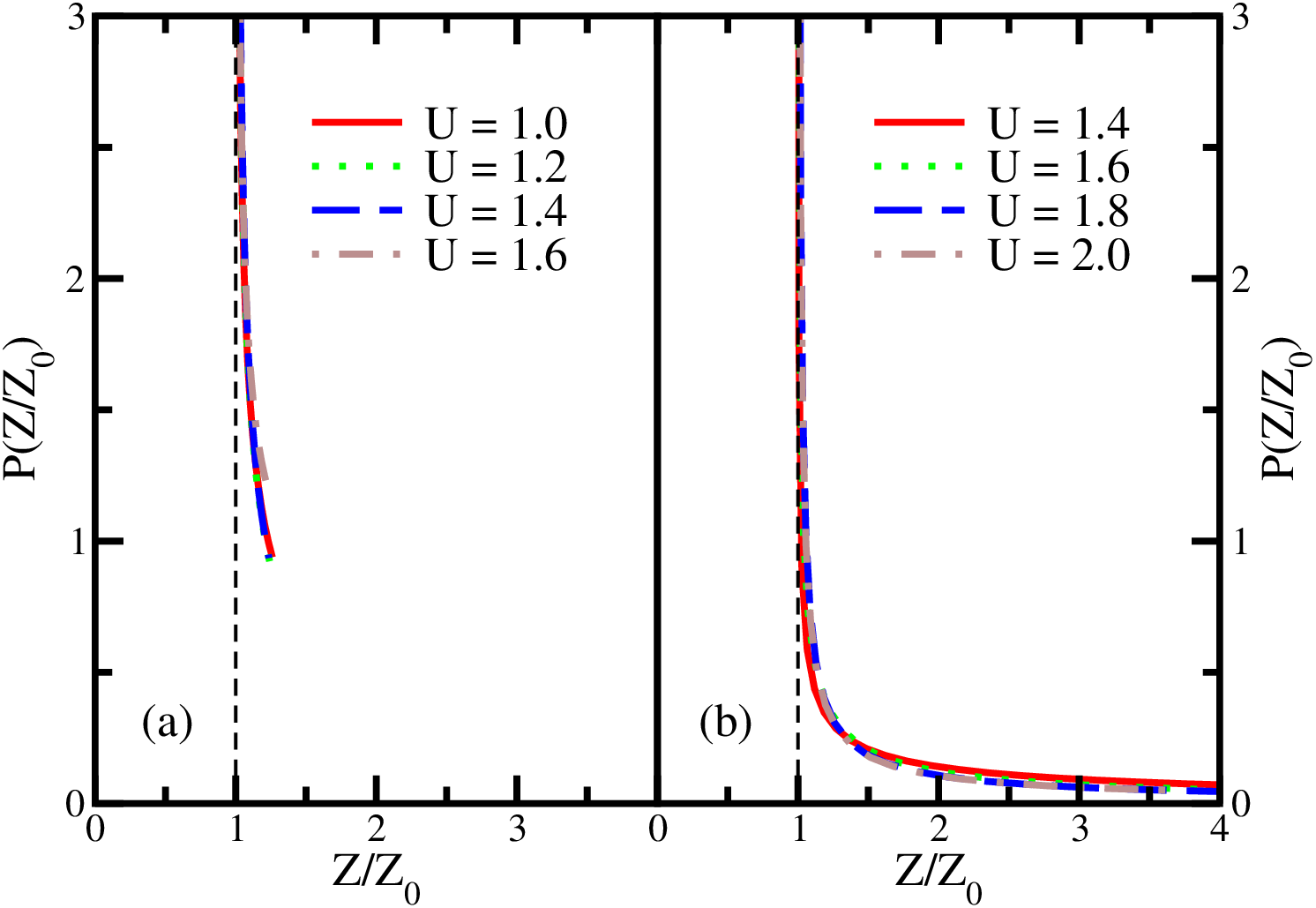}
\end{center}
\caption{Distribution of $Z/Z_{0}$ for (a) smaller ($W=0.5$) and (b) larger
($W=1.5$) disorder than the one in Fig.~\ref{fig4}. Other parameters used was
$T=0.005$.}%
\label{fig5}%
\end{figure}

Equivalently, we can characterize the QP weights by their probability
distribution function $P(Z_{i})$. Typical results for $P(Z_{i})$ are shown in
the inset of Fig.~\ref{fig4}. As the $Z_{i}$ decrease near the transition, 
the distribution function $P(Z_{i})$ changes its form and narrows down.
However, if our scaling hypothesis is valid, then the \textit{shape} of this
distribution should approach a ``fixed-point" form very close to the
transition. More precisely, we expect the distribution for reduced QP weights
$P(Z_{i}^{\ast})$ to collapse to a single scaling function close to $U_{c2}$. Results confirming precisely such behavior are presented in Fig.~\ref{fig4}.

An interesting question relates to the precise form of the fixed-point
distribution function $P(Z_{i}^{\ast})$, and how it may depend on disorder. In
the clean limit, obviously, it reduces to $\delta(Z_{i}^{\ast}-1)$ indicating
that spatial fluctuations are suppressed. As the disorder increases,
$P(Z_{i}^{\ast})$ becomes very broad (as shown in Fig.~\ref{fig5}), reflecting large
site-to-site fluctuations in the local QP weights. This behavior may be
regarded as a precursor of electronic Griffiths phases,\cite{griffiths} which
emerge for stronger disorder, as found within \textit{stat}DMFT
approaches.\cite{statdmft}

In essential contrast to the clean limit, the approach to the Mott transition
in presence of disorder thus needs to be characterized by the entire
\textit{probability distribution function} of QP parameters. At first glance,
this may appear to require a description considerably more complex than in
the absence of disorder. However, we have demonstrated that in the critical region
the distributions approach a fixed point form, allowing for \textquotedblleft
single parameter scaling," in close analogy to the clean Mott transition. This
finding immediately suggests that our arguments describing the finite
temperature coexistence behavior in the clean limit may successfully be
extended to the disordered case as well, allowing for a complete qualitative
description, which we discuss in the following section.

\section{Coexistence region in presence of disorder}

Within the DMFT formulation, the disorder is not expected to qualitatively
affect the temperature dependence of the insulating spinodal, since the forms
of the Hubbard bands remain qualitatively similar to that in the clean limit.
The principal effect of disorder in the Mott insulating phase is to simply
broaden the Hubbard bands, which retain well defined (sharp) band edges due to
the CPA-like treatment of randomness in the DMFT limit. Indeed, our
quantitative results [see Fig.~\ref{fig1}(a)] confirm that 
$U_{c1}(T)\approx U_{c1}(0)$
retains very weak temperature dependence, as in the clean case. The only
modification is that $U_{c1}(0)$ rapidly grows as disorder is increase,
reflecting the disorder-induced broadening of the Hubbard bands.

\begin{figure}[ptb]
\begin{center}
\includegraphics[
trim=0.000000in 0.000000in 0.000000in -1.1in,
height=2.2in, width=2.6in ]{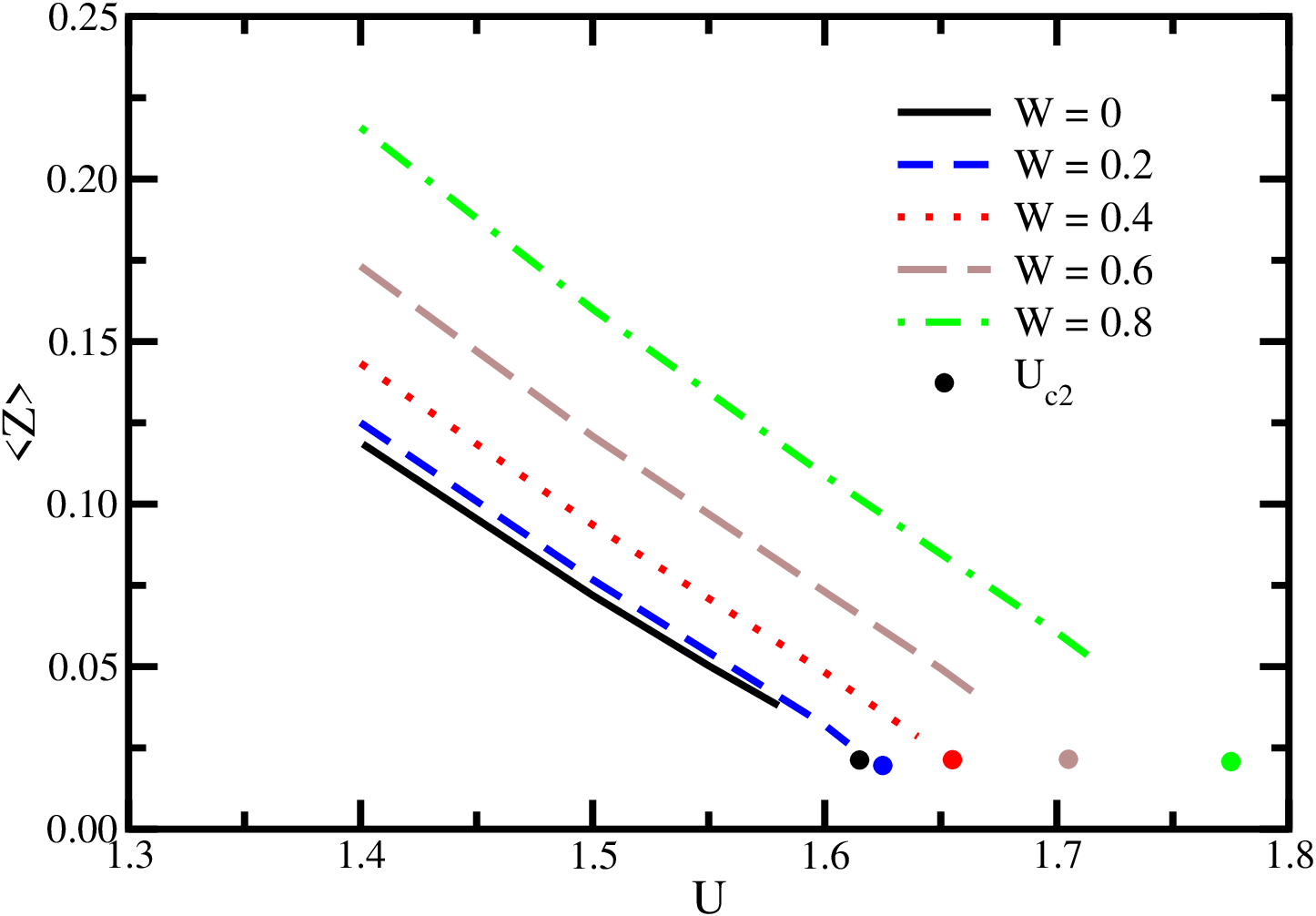}
\end{center}
\caption{Average quasiparticle weight as a function of the interaction
potential for different values of disorder.}%
\label{fig6}
\end{figure}

The metallic solution is again found to be unstable above a certain coherence
temperature $T^{\ast}(W,U)$, which defines the locus of the metallic spinodal
$U_{c2}(T)$. An added subtlety is that different sites start to decohere at
different temperatures, an effect that earlier work\cite{inelastic} found
responsible for a nearly-linear temperature dependence of the resistivity in
the disordered metallic phase. Nevertheless, sufficiently close to the Mott
transition (within the coexistence region), a sharply defined temperature
scale $T^{\ast}(W,U)$ emerges where the metallic solution suddenly disappears
and where the qualitative form of the spectrum changes on \textit{all }sites.
This temperature scale defines the locus of the metallic spinodal,
corresponding to the equation
\begin{equation}
T=T^{\ast}(W,U_{c2}).
\end{equation}

\begin{figure}[ptb]
\begin{center}
\includegraphics[
trim=0.000000in 0.000000in 0.000000in -1.1in,
height=2.2in, width=2.6in ]{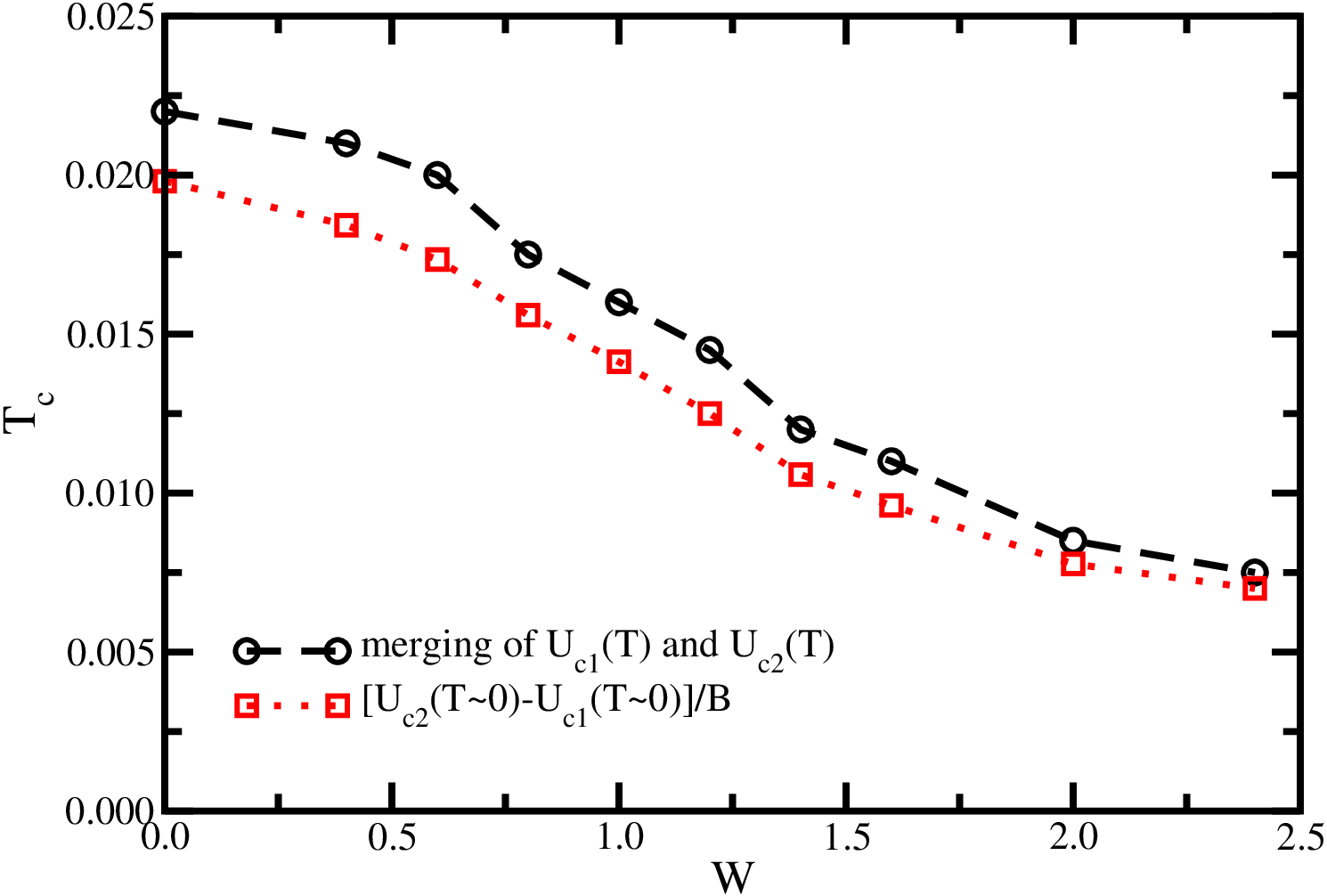}
\end{center}
\caption{Temperature at which the $U_{c1}$ and $U_{c2}$ lines merge in the
$(U,T)$ phase diagram as a function of disorder. The plot shows both the
results obtained directly from the numerical data as well as those calculated
from the linear fitting to the $U_{c2}(T)$ line. In the latter, $T_{c}$ was
calculated using the values of $U_{c1}$ at $T=0.005$, except for $W=0$ where
the result at $T=0.0075$ was used.}%
\label{fig7}
\end{figure}

At first glance, it is anything but obvious how $T^{\ast}(W,U)$ should be
estimated. As in the clean case, the reduction of this temperature scale as
the transition is approached must reflect the behavior of the local
quasi-particle weights $Z_{i}$, and presumably depend on the precise form of
the distribution function $P(Z_{i})$. As we have seen, however, all the local
QP weight scale in a similar fashion in the critical regime, which suggests that a
reasonable estimate may be obtained simply from their average value
\begin{equation}
\left\langle Z_{i}\right\rangle =\int d\varepsilon_{i}P(\varepsilon_{i})Z_{i}.
\end{equation}
At least for sufficiently weak disorder, we may expect that [cf. Eq.\ (5)]%
\begin{equation}
T^{\ast}(W,U)\approx AT_{F}\left\langle Z_{i}\right\rangle ,
\end{equation}
where $A\approx0.2$ as in the clean case. Using the fact that all $Z_{i}$'s
decrease linearly near the transition, we expect%
\begin{equation}
\left\langle Z_{i}\right\rangle =C(W)[U_{c2}(0)-U].
\end{equation}
To confirm this, we explicitly calculated $\left\langle Z_{i}\right\rangle $
as a function of $U$ for different levels of disorder; the results are shown in Fig.~\ref{fig6}. We
conclude that $C(W)\approx C(0)\approx0.45$. These results suggest that the
metallic spinodal should take the form,
\begin{equation}
U_{c2}(W,T)\approx U_{c2}(W,0)-B(W)T,
\end{equation}
where $B(W)\approx B(0)=22$. Our non-zero temperature results for $U_{c2}(W,T)$
[see Fig.~\ref{fig1}(a)] fully confirm these expectations. Based on these results, we
finally obtain the desired expression for $T_{c}(W)$ of the form
\begin{equation}
T_{c}(W)\approx\lbrack U_{c2}(W,0)-U_{c1}(W,0)]/B(0).
\end{equation}
To test the proposed procedure, we have used the values for $U_{c1}(W)$ and
$U_{c2}(W)$ at the lowest temperature of our calculation ($T=0.005$) to
estimate $T_{c}(W)$. As we can see from Fig.~\ref{fig7}, our analytical estimates are
found to be in excellent agreement with results of explicit non-zero temperature
calculations. The decrease of $T_{c}(W)$ with disorder thus directly reflects
the \textquotedblleft shrinking" of the coexistence region at low temperature,
which in its turn reflects the decrease of the energy difference between the
metallic and the insulating solution.

\section{Conclusions}

In this paper we have used a DMFT approach to examine the effects of disorder
on the critical behavior near the Mott metal-insulator transition, with
special emphasis on non-zero temperature properties associated with the two 
spinodal lines $U_{c1}$ and $U_{c2}$. By using a combination of numerical 
results and analytical arguments we have demonstrated that simple scaling 
behavior emerges, providing a complete description of the critical regime. 

In contrast to the clean case, the presence of disorder requires one to 
examine the entire distribution of local spectral functions, $\rho_i (\omega)$,
describing how the local spectra varies with position in the sample. This 
can be probed with scanning tunneling microscopy (STM). Notice that the 
distribution function describing the site dependence of $\rho_i (\omega)$ 
will depend on the frequency of observation: it will be broader at higher 
energies [as seen in Fig.~\ref{fig2}(b)], where
a real space picture is appropriate to describe the Hubbard bands, and
narrower at low frequencies, where a quasiparticle description in k space is 
appropriate. This is a manifestation of frequency dependence of the 
disorder screening discussed in an earlier paper by some of 
us.\cite{screening} 

In the metallic regime, at low temperatures, the spectral function can be 
parametrized in terms of the distribution of quasiparticle parameters, which  
displays simple scaling properties. This allowed us to characterize the 
behavior near $U_{c2}$ using a single parameter scaling procedure. The 
approach to $U_{c2}$ thus retains a character qualitatively independent of 
the level of disorder, where the vanishing of quasiparticle weight signals 
the transmutation of itinerant electrons into localized magnetic moments. 

Within the examined DMFT formulation, the region between the two spinodal 
lines $U_{c1}$ and $U_{c2}$ although reduced in size and extent cannot be 
completely eliminated no matter how large the disorder. Of course, these 
predictions are applicable only for weak enough disorder where Anderson 
localization effects can be ignored. Extensions of DMFT that incorporate 
Anderson localization mechanisms at zero temperature are 
available,\cite{statdmft,vollhardt} but applying these approaches to examine 
the non-zero temperature behavior near Mott-Anderson transitions remains an 
interesting research direction. The behavior at the first order transition line and 
the actual nucleation of either the metallic or insulating phase, between 
$U_{c1}$ and $U_{c2}$, are also strongly modified by disorder, and this as
well is left for future study.

The authors thank A. Georges and D. Tanaskovi\'{c} for useful discussions.
This work was supported by NSF grants DMR-9974311 and DMR-0234215 (VD) and
DMR-0096462 (GK).


\begin{thebibliography}{99}
\bibitem {coeffA}The definition of the coherence temperature $T^{\ast}$ used
in Ref. [9] was based on the temperature dependence of the resistivity. It was
defined as the temperature where deviations from the low temperature $T^{2}$
form became appreciable, and as such represents a crossover scale. That work
found $A\approx0.1$, which is in reasonable agreement with our present result
$A\approx0.2$, given the ambiguity in precisely defining the resistivity
crossover scale.
\bibitem {sitener}In addition to the quasiparticle weight, the low energy
excitations are also characterized by renormalized site energies describing
the renormalized disorder potential. However, as discussed in Ref. [8], within
the DMFT scheme the renormalized disorder is strongly screened, so the local
quasiparticle weights $Z(\varepsilon_{i})$ are sufficient to characterize the
critical behavior.
\end{thebibliography}
\end{document}